\documentclass[12pt,thmsa]{article}
\usepackage{amssymb}
\usepackage{epsfig}

\input{tcilatex}

\begin{document}

\author{John D. Barrow\thanks{%
e-mail: \texttt{J.D.Barrow@damtp.cam.ac.uk}} \ and \ David F. Mota\thanks{%
e-mail: \texttt{D.F.Mota@damtp.cam.ac.uk}} \\
{\normalsize \textsl{Department of Applied Mathematics and Theoretical
Physics}}\\
{\normalsize \textsl{Centre for Mathematical Sciences, University of
Cambridge}}\\
{\normalsize \textsl{Wilberforce Road, Cambridge CB3 0WA, UK}}}
\title{Gauge-Invariant Perturbations of Varying-Alpha Cosmologies}
\date{}
\maketitle

\begin{abstract}
Using a gauge-invariant formalism we derive and solve the perturbed
cosmological equations for the BSBM theory of varying fine structure
'constant'. We calculate the time evolution of inhomogeneous perturbations
of the fine structure constant, $\frac{\delta \alpha }{\alpha }$ on small
and large scales with respect to the Hubble radius. In a radiation-dominated
universe small inhomogeneities in $\alpha $ will decrease on large scales
but on scales smaller than the Hubble radius they will undergo stable
oscillations. In a dust-dominated universe small inhomogeneous perturbations
in $\alpha $ will become constant on large scales and on small scales they
will increase as $t^{2/3}$, and $\frac{\delta \alpha }{\alpha }$ will track $%
\frac{\delta \rho _{m}}{\rho _{m}}$ . If the expansion accelerates, as in
the case of a $\Lambda $ or quintessence-dominated phase, inhomogeneities in 
$\alpha $ will decrease on both large and small scales. The amplitude of
perturbations in $\alpha $ will be much smaller than that of matter or
radiation perturbations. We also present a numerical study of the non-linear
evolution of spherical inhomogeneities in radiation and dust universes by
means of a comparison between the evolution of flat and closed Friedmann
models with time-varying $\alpha .$Various limitations of these simple
models are also discussed.
\end{abstract}


\section{Introduction}

A third sample of observations of quasar absorption-line spectra has been
found to be consistent with a time variation in the value of the fine
structure 'constant' between redshifts $z=0.2-3.7$ and the present. The
entire data set of 128 objects gives spectra consistent with a shift of $%
\Delta \alpha /\alpha \equiv \lbrack \alpha (z)-\alpha _{0}]/\alpha
_{0}=-0.57\pm 0.10\times 10^{-5},$ where $\alpha _{0}$ is the present value
of the fine structure constant \cite{murphy,webb2,webb,webb3}. Extensive
analysis has yet to find a selection effect that can explain the sense and
magnitude of the relativistic line-shifts underpinning these deductions.
Motivated by these observations, there has been considerable theoretical
investigation of the cosmological consequences of varying $\alpha $, \cite%
{book}$.$ Barrow, Sandvik, and Magueijo \cite%
{bsbm,bsm1,bsm2,bsm3,Barrow:2002ed} have studied in detail the cosmological
consequence of an extension of the varying-$\alpha $ Maxwell equations
formulated by Bekenstein \cite{bek2}. We call these BSBM theories. They
allow us to understand the effects of the expansion of the universe on
variations of $\alpha $ and to evaluate the effects of varying $\alpha $ on
free fall which leads to potentially observable violations of the weak
equivalence principle \cite{bsm4,zal,olive,poly}. They also allow us to
investigate whether or not other cosmological observations are consistent
with the small variations in $\alpha $ that are required to fit the quasar
observations. Other theories, founded explicitly upon variation in the
velocity of light, have also been proposed but their main interest is in
connection with events in the very early universe \cite{moffatal,am,ba,mof}
and the problem of whether or not non-inflationary explanations for
particular features of the large-scale structure of the universe are
possible \cite{jbvsl}. Others have begun to consider the implications
for grand unification of coupling strengths, \cite%
{mar,poly,jdb,drink,banks,guts,cal,arm,chacko,correia}, and
astronomical probes of the
constancy of the electron-proton mass ratio have reported possible evidence for
time variation, \cite{ivan}, but as yet the statistical significance is low.

Barrow, Sandvik, and Magueijo \cite{bsbm,bsm1,bsm2,bsm3} have shown that
BSBM theories have a number of appealing properties. They predict that there
should be no variation of $\alpha $ during the radiation era in our universe
and none during any present or late-time curvature or cosmological constant
dominated era. During the dust era $\alpha $ should grow (leading to $\Delta
\alpha /\alpha <0$ as observed) but only as $\log (t).$This behaviour allows
the quasar data to be accommodated without producing conflict with recent
geonuclear limits on allowed variations of $\alpha $, like the Oklo natural
reactor limits of $\Delta \alpha /\alpha \lesssim 10^{-7}$ \cite{shly,
sis,lan,dd, fuj} because they are imposed at a very low effective redshift
of $z\approx 0.15,$ at which time the universe has begun accelerating and
all variations in $\alpha $ are damped out. Recent deductions of a possible
upper limit of $\Delta \alpha /\alpha \lesssim 3\times 10^{-7}$ at $z=0.45$
from nuclear $\beta $ decays are potentially more restrictive \cite{pd,dys,
olive1}. However, it must be remembered that both these nuclear limits are
derived from a local solar-system environment. In the absence of a theory
relating the value of, and rate of change of, $\alpha $ on the cosmological
scales where quasar lines form to their values in the virialised local
inhomogeneities where galaxies, stars and planets form, one should be wary
of ignoring the possible corrections that must be introduced when comparing
planetary and quasar bounds: the density of the Earth would not, for
example, be a good indicator of the density of the universe. Thus,
inhomogeneity is an important factor in the study of varying-constant
cosmological theories. In this paper we are going to study the evolution of
small, gauge-invariant perturbations to the exact Friedmann-Robertson-Walker
solutions of the BSBM theory. The results of this investigation will reveal
whether one must worry about fast growth of small initial inhomogeneities in
the value of $\alpha $, which would lead to spatial variations in the value
of $\alpha $ that might be more significant than the time variations at late
times.

Bekenstein \cite{bek2} generalised Maxwell's equations to include varying $e$
and this theory was then generalised by Sandvik, Barrow and Magueijo \cite%
{bsbm} to include gravitation. In the BSBM varying-$\alpha $ theory, the
quantities $c$ and $\hbar $ are taken to be constant, while $e$ varies as a
function of a real scalar field $\psi ,$ with $\alpha =e^{2},$ hence 
\begin{eqnarray}
e &=&e_{0}\exp [\psi ]  \label{e} \\
\alpha &=&\alpha _{0}\exp [2\psi ]
\end{eqnarray}

Our discussion is organised as follows. In section 2 of the paper we give
the gauge-invariant perturbation equations following the development used in
general relativity. In section 3 we specify the BSBM cosmology with varying $%
\alpha $ and derive the gauge invariant linear perturbation equations which
couple the perturbations in the gravitational field to those in $\alpha $
and the density. In section 4 we solve for the time-evolution of small
inhomogeneities in the fine structure 'constant' on large and small scales
for radiation, dust, and cosmological constant dominated expansion of the
background universe. In section 5 we extend these studies into the
non-linear regime by means of numerical solutions for flat and closed FRW
universes. The evolution of spherical curvature inhomogeneities in density
and in $\alpha $ is followed by computing the difference in time evolution
between the FRW models of different curvature. These studies also reveal for
the first time the behaviour of $\alpha $ in closed FRW models in the BSBM
theory. These results are discussed and conclusions drawn in section 6.

Units will be used in which $c=\hbar =1$; Greek indices run form $0$ to $3$
and Latin indices only over the spatial degrees of freedom $1$ to $3$. The
Einstein summation convention is assumed; $a(t)$ is the scale factor of the
background Friedmann-Robertson-Walker (FRW) metric and $G$ is Newton's
gravitation constant.

\section{The Background and Perturbations}

In what follows we shall assume that the metric of space-time deviates only
by a small amount from a homogeneous, isotropic FRW space-time which is
defined to be the background universe. In this case, it is convenient to
split the metric into two parts: the background metric, and its
perturbation. We observe that the universe is nearly homogeneous and
isotropic on large scales (metric perturbations to the cosmic microwave
background radiation are observed to be small $\approx 10^{-5}$ $<<1$).

The background line element is:

\begin{equation}
ds^{2}=^{(0)}g_{\mu \nu }dx^{\mu }dx^{\nu }=dt^{2}-a^{2}(t)\gamma
_{ij}dx^{i}dx^{j}=a^{2}(\eta )(d\eta ^{2}-\gamma _{ij}dx^{i}dx^{j}),
\label{backline}
\end{equation}%
where $\eta $ is the conformal time,

\[
d\eta =a^{-1}dt. 
\]

We choose the background metric to be the FRW metric, so

\[
\gamma _{ij}=\delta _{ij}[1+\kappa (x^{2}+y^{2}+z^{2})]^{-2}, 
\]%
where $\kappa =0,1,-1$ depending on whether the three-dimensional
hypersurfaces of constant $\eta $ are flat, closed or open. The Einstein
equations are:

\begin{equation}
G_{\nu }^{\mu }=R_{\nu }^{\mu }-\frac{1}{2}\delta _{\nu }^{\mu }R=8\pi
GT_{\nu }^{\mu },
\end{equation}%
where $R_{\nu }^{\mu }$ is the Ricci tensor, $R\equiv R_{\mu }^{\mu }$ is
the Ricci scalar and $T_{\nu }^{\mu }$ is the total energy-momentum tensor.
We shall exploit the fact that the BSBM theory for varying $\alpha $ can be
expressed as general relativity with a particular linear combination of
energy-momentum tensors. For the moment consider the presence of a single
energy-momentum tensor. In the next section this will be decomposed
appropriately.

For the background metric in equation (\ref{backline}), in conformal time,
the Einstein equations reduce to the $0-0$ equation

\begin{equation}
\mathcal{H}^{2}=\frac{8\pi G}{3}a^{2}T_{0}^{0}-\kappa ,  \label{00}
\end{equation}%
where $\mathcal{H}\equiv \frac{a^{\prime }}{a}$ using conformal time, and
the $i-i$ equation:

\begin{equation}  \label{ii}
\mathcal{H^{\prime}}+ \mathcal{H}^2 = \frac{4\pi G}{3}a^2 T - \kappa, \qquad
T\equiv T^{\mu}_{\mu}
\end{equation}

For the background metric (\ref{backline}), the space-space part of the
Ricci tensor $R_{j}^{i}$ is proportional to $\delta _{j}^{i}$. Thus, for an
isotropic background universe, the energy-momentum tensor must also be
spatially diagonal, $T_{j}^{i}\propto \delta _{j}^{i}$, in order that the
Einstein equations are satisfied. Differentiating (\ref{00}) with respect to 
$\eta $ and subtracting $2a^{\prime }$ we get the continuity equation for
matter $\triangledown _{\mu }T_{0}^{\mu }=0$ :

\begin{equation}  \label{conserv}
dT^{0}_{0}=-(4T^{0}_{0}-T)d\ln{a}.
\end{equation}

We now introduce small perturbations around the FRW background and follow
the gauge-invariant approach of \ Mukhanov \cite{mukhanov}. The full line
element may be expressed as:

\begin{eqnarray}  \label{fullline}
ds^2 &=& ^{(0)}g_{\mu\nu} dx^{\mu}dx^{\nu} + \delta g_{\mu\nu}
dx^{\mu}dx^{\nu},
\end{eqnarray}
where $\delta g_{\mu\nu}$ describes the perturbation. The full metric has
been decoupled into its background parts and perturbation parts:

\begin{eqnarray}  \label{metric}
g_{\mu\nu} &=& ^{(0)}g_{\mu\nu}+ \delta g_{\mu\nu}.
\end{eqnarray}

In general, the metric perturbations can be divided into three distinct
types: scalar, vector and tensor. Neither of the vector and tensor
perturbations exhibit growing instabilities in dust and radiation universes
. Vector perturbations decay kinematically in an expanding universe whereas
tensor perturbations lead to gravitational waves that do not couple to the
isotropic energy-density and pressure inhomogeneities. However, scalar
perturbations may lead to growing inhomogeneities which, in turn, have an
important effect on the dynamics of matter and thereby on the time and space
variations of the fine structure constant in the BSBM\ theory. In the linear
approximation, scalar, vector and tensor perturbations evolve independently
and can be considered separately. In this paper we will consider only the
scalar perturbation modes.

The most general form of the scalar metric perturbations is constructed
using four scalar quantities which are functions of space and time
coordinates:

$\delta g_{\mu \nu }=%
\bordermatrix{ & & \cr
		 & 2\phi  & -B_{|i}   \cr
                 & -B_{|i} & 2(\chi \gamma_{ij}-E_{|ij}) \cr},$

where $|i$ represents the three-dimensional covariant derivative.

From the above equation and eqn. (\ref{metric}), the line element for the
background and for the scalar metric perturbations is

\begin{equation}
ds^{2}=a^{2}(\eta )\{(1+2\phi )d\eta ^{2}-2B_{|i}dx^{i}d\eta -[(1-2\chi
)\gamma _{ij}+2E_{|ij}]dx^{i}dx^{j})\}.  \label{perturbedline}
\end{equation}

\subsection{Gauge-invariant Variables}

Gauge-invariant variables are unchanged under all infinitesimal scalar
coordinate transformations, so they are independent of the background
coordinates. Such quantities can be constructed out of the four scalar
functions $\Phi $, $\Psi $, $E$ and $B$ \cite{bardeen}. The simplest
gauge-invariant linear combinations which span the space of gauge-invariant
variables that can be constructed from the metric variables alone are:

\begin{equation}
\Phi =\phi +\frac{[(B-E^{\prime })a]^{\prime }}{a},\qquad \Psi =\chi -\frac{a%
}{a^{\prime }}\left( B-E^{\prime }\right) .  \label{gauge}
\end{equation}

In general, a scalar quantity $f(\eta ,\mathbf{{x})}$ defined in the
spacetime can be split into its background value and a perturbation $f(\eta ,%
\mathbf{x})=f_{0}(\eta )+\delta f(\eta ,\mathbf{x})$. Since, in general, $%
\delta f$ is not gauge invariant, we cannot use this scalar quantity without
modification if we want to have gauge-invariant equations. Hence, we
consider the following gauge-invariant combination:

\begin{equation}
\delta f^{(gi)}=\delta f+f_{0}^{\prime }(B-E^{\prime }).  \label{gaugeinv}
\end{equation}

The freedom of gauge choice can be used to impose two conditions on the four
scalar functions. The \textit{longitudinal gauge} is defined by the
conditions $E=B=0$. This gauge choice has the advantage of ruling out the
complications of residual gauge modes. Also, in this gauge $\phi $ and $\chi 
$ coincide with the gauge-invariant variables $\Phi $ and $\Psi $
respectively. In this longitudinal gauge, the metric takes the form:

\[
ds^{2}=a^{2}(\eta )[(1+2\Phi )d\eta ^{2}-(1-2\Psi )\gamma
_{ij}dx^{i}dx^{j})], 
\]%
and the gauge invariants $\Phi $ and $\Psi $ become the amplitudes of the
metric perturbations in the longitudinal coordinate system. In the case
where there are no space-space components in the energy-momentum tensor, so $%
T_{j}^{i}\propto \delta _{j}^{i}$, we have that $\Phi =\Psi $ and there
remains only one free metric perturbation variable which is a generalisation
of the Newtonian gravitational potential.

\section{Linear Theory for Cosmological Perturbations}

For small perturbations of the metric, the Einstein tensor can be written as 
$G_{\nu }^{\mu }=\ ^{(0)}G_{\nu }^{\mu }+\delta G_{\nu }^{\mu }$, and the
energy-momentum tensor can be split in a similar way. The fully perturbed
Einstein equations can be obtained for scalar perturbation modes with the
line element given by (\ref{perturbedline}). However, these equations are
not gauge invariant, since they contain non gauge-invariant quantities. In
order to have gauge-invariant equations we need to replace $\phi $ and $\chi 
$ by the gauge-invariant variables $\Phi $, $\Psi $ and $B-E^{\prime }$ and
to construct the gauge-invariant equivalents of $\delta G_{\nu }^{\mu }$ and 
$\delta T_{\nu }^{\mu }$ we need to rewrite then as \cite{mukhanov}:

\begin{eqnarray}
\delta G_{0}^{(gi)\ 0}=\delta G_{0}^{0}+(^{(0)}G_{0}^{0})^{\prime
}(B-E^{\prime }), \\
\delta G_{j}^{(gi)\ i}=\delta G_{j}^{i}+(^{(0)}G_{j}^{i})^{\prime
}(B-E^{\prime }),  \nonumber \\
\delta G_{i}^{(gi)\ 0}=\delta G_{i}^{0}+(^{(0)}G_{0}^{0}-\frac{1}{3}\
^{(0)}G_{k}^{k})(B-E^{\prime })_{|i},  \nonumber
\end{eqnarray}
and analogously for $\delta T_{\nu }^{\mu }$, 
\[
\delta T_{0}^{(gi)\ 0}=\delta T_{0}^{0}+(^{(0)}T_{0}^{0})^{\prime
}(B-E^{\prime }), 
\]%
\[
\delta T_{j}^{(gi)\ i}=\delta T_{j}^{i}+(^{(0)}T_{j}^{i})^{\prime
}(B-E^{\prime }), 
\]%
\[
\delta T_{i}^{(gi)\ 0}=\delta T_{i}^{0}+(^{(0)}T_{0}^{0}-\frac{1}{3}\
^{(0)}T_{k}^{k})(B-E^{\prime })_{|i}. 
\]

The components of the perturbed Einstein equations linearised around small
perturbations of the background are $\delta G_{\nu }^{(gi)\ \mu }=8\pi
G\delta T_{\nu }^{(gi)\ \mu }$:

\begin{eqnarray}  \label{genpertein}
\delta G_{0}^{0}=\triangledown ^{2}\Phi -3\mathcal{H}\Phi ^{\prime }-3\Phi (%
\mathcal{H}^{2}-\kappa )=4\pi Ga^{2}\delta T_{0}^{(gi)0}  \nonumber \\
\delta G_{i}^{0}=\partial _{i}(a\Phi )^{^{\prime }}=4\pi Ga\delta
T_{i}^{(gi)0} \\
\delta G_{i}^{j}=\Phi ^{\prime \prime }+3\mathcal{H}\Phi ^{\prime }+\Phi (2%
\mathcal{H}^{\prime }+\mathcal{H}^{2}-\kappa )=-4\pi Ga^{2}\delta
T_{j}^{(gi)i}  \nonumber
\end{eqnarray}
where we have already simplified the equations since the energy-momentum
tensor that we will be considering in section 3 has no space-space
components and so $\Phi =\Psi $.

In order to close our system of equations, we need equations of motion for
the matter formulated in a gauge-invariant way. This requires explicit
expressions for the energy-momentum tensor and so we must now specify the
BSBM theory.

\subsection{The Model and the Background Equations}

\subsubsection{The BSBM Theory}

The action for the universe in the BSBM theory is given by:

\begin{equation}
S=\int d^{4}x\sqrt{-g}\left( \mathcal{L}_{grav}+\mathcal{L}_{matter}+%
\mathcal{L}_{\psi }+\mathcal{L}_{em}e^{-2\psi }\right) ,  \label{S}
\end{equation}%
where $\mathcal{L}_{\psi }={\frac{\omega }{2}}\partial _{\mu }\psi \partial
^{\mu }\psi $, $\omega $ is a coupling constant, $\mathcal{L}_{em}=-\frac{1}{%
4}f_{\mu \nu }f^{\mu \nu },$ and $\psi $ was defined in eqn. (\ref{e}). The
gravitational Lagrangian is the usual $\mathcal{L}_{g}=-\frac{1}{16\pi G}R$,
with $R$ the curvature scalar, and we have defined an auxiliary gauge
potential $a_{\mu }=\epsilon A_{\mu },$ where $\varepsilon (x^{\mu
})=e/e_{0} $ describes change in the electron charge away from a constant
reference value $e_{0}$. The field tensor $f_{\mu \nu }=\epsilon F_{\mu \nu
}=\partial _{\mu }a_{\nu }-\partial _{\nu }a_{\mu }$, so the covariant
derivative takes the usual form, $D_{\mu }=\partial _{\mu }+ie_{0}a_{\mu }$.
The dependence on $\epsilon $ in the Lagrangian then occurs only in the
kinetic term for $\epsilon $ and through the $F^{2}=f^{2}/\epsilon ^{2}$
term.

It was shown in \cite{Barrow:2002ed} and \cite{bsm3} that, in the context of
the BSBM model, the homogeneous evolution of $\psi $ does not create
significant metric perturbations at late times and the cosmological
time-evolution of the expansion scale-factor is very well approximated by
the usual power-laws found in $\kappa =0$ FRW models filled with a perfect
fluid.

Therefore, we will assume there are no major modifications in the perturbed
spacetime which would lead to changes of the behaviour of the perturbed
variables of a perturbed FRW spacetime filled with perfect fluid. That is,
we will assume that the energy-density perturbations and the metric
potential are the same as in a FRW universe with no variation of $\alpha $.
Physically, this is to be expected for most of the evolution, although this
assumption might break down (along with much else) on approach to initial
and final cosmological singularities. It is a reflection of the fact that
the changes in $\alpha $ have negligible feedback into the changes in the
expansion, which are governed to leading order by gravity. The principal
effects are those of perturbations in the matter density and expansion rate
on the evolution of $\alpha .$ This simplification will allow us to write
the time and space variations of the scalar field, $\delta \psi $, as a
functions of $\rho _{m}$, $\rho _{r}$, $\delta \rho _{m}$, $\delta \rho _{r}$%
, $\Phi $ and $\psi $, where $\delta \rho _{m}$, $\delta \rho _{r}$ and $%
\Phi $ will be given by the solutions found in ref. \cite{mukhanov} for
universes with no variation of $\alpha ;$ the field $\psi $ will be given by
the solutions found previously in \cite{Barrow:2002ed}. In order to find an
expression for $\delta \psi $ in terms of these quantities we need to write
the gauge-invariant linearly perturbed Einstein equations for the BSBM model.

\subsubsection{The Background Equations}

We vary the action (\ref{S}) with respect to the metric to obtain the
generalised Einstein equations:

\[
G_{\nu }^{\ \mu }=8\pi G\left( T_{\nu }^{\mathit{matter}\ \mu }+T_{\nu
}^{\psi \ \mu }+T_{\nu }^{\mathit{em}\ \mu }\right) 
\]%
where $T_{\mu \nu }^{\mathrm{mat}}=\frac{2}{\sqrt{-g}}\frac{\delta (\sqrt{-g}%
\mathcal{L}_{\mathrm{mat}})}{\delta g^{\mu \nu }}$ is the energy-momentum
tensor for perfect-fluid matter fields, and

\[
T_{\mu \nu }^{\mathrm{matter}}=(\rho _{m}+p_{m})u_{\mu }u_{\nu }-p_{m}g_{\mu
\nu }, 
\]%
where $u^{\mu }$ $=\delta _{0}^{\mu }$ is the comoving fluid 4-velocity; $%
T_{\nu }^{\psi \ \mu }$ and $T_{\nu }^{\mathit{em}\ \mu }$ are the
energy-momentum tensors for the kinetic energy of the field $\psi $ and the
electromagnetic field respectively:

\[
T_{\mu \nu }^{\psi }=\omega \partial _{\mu }\psi \partial _{\nu }\psi -{%
\frac{\omega }{2}}g_{\mu \nu }\partial _{\beta }\psi \partial ^{\beta }\psi
,\quad T_{\mu \nu }^{\mathrm{em}}=F_{\mu \beta }F_{\nu }^{\beta }e^{-2\psi }-%
{\frac{1}{4}}g_{\mu \nu }F_{\sigma \beta }F^{\sigma \beta }e^{-2\psi }. 
\]%
Note, the total energy density of the electromagnetic field is the sum of
the Coulomb energy density $\zeta \rho _{m}$ and the radiation energy
density $\rho _{r}$ , where $-1\leq \zeta \leq 1$ is the fraction of mass
density $\rho _{m}$ of matter in the form of the Coulomb energy. We will
then consider $T_{\nu }^{\mathit{em}\ \mu }$ as a perfect fluid: 
\[
T_{\mu \nu }^{\mathrm{em}}=(|\zeta |\rho _{m}+\rho
_{r}+p_{m}+p_{r})e^{-2\psi }u_{\mu }u_{\nu }-\left( p_{m}+p_{r}\right)
e^{-2\psi }g_{\mu \nu } 
\]

The propagation equation for $\psi $ comes from the variational principle as:

\begin{equation}
{\partial _{\mu }\left[ \sqrt{-g}g^{\mu \nu }\partial _{\nu }\psi \right] =-%
\frac{2}{\omega }}\sqrt{-g}e^{-2\psi }\mathcal{L}_{\mathrm{em}}.
\label{psieq}
\end{equation}%
This equation determines how $e,$ and hence $\alpha $, varies with time. It
is clear that $\mathcal{L}_{em}$ vanishes for a sea of pure radiation
because $\mathcal{L}_{em}=(E^{2}-B^{2})/2=0$. In order to make quantitative
predictions we need to know how non-relativistic matter contributes to the
right hand side for equation (\ref{psieq}), through $\mathcal{L}_{em}=\zeta
\rho _{m}$.

The background equations can now be explicitly obtained:

\begin{eqnarray}
3{{\mathcal{H}}^{2}} &=&8\pi Ga^{2}\left( \rho _{m}+\left( \rho _{r}+|\zeta
|\rho _{m}\right) e^{-2\psi }+\frac{\omega }{2}a^{-2}{{\psi ^{\prime }}^{2}}
\right) +a^{2}\Lambda -3\kappa  \label{fried1} \\
3{\mathcal{H^{\prime }}} &=&-4\pi Ga^{2}\left( \rho _{m}\left( 1+|\zeta
|e^{-2\psi }\right) +2\rho _{r}e^{-2\psi }+2\omega a^{-2}{{\psi ^{\prime }}%
^{2}}\right) +a^{2}\Lambda  \nonumber
\end{eqnarray}%
where $\Lambda $ is the cosmological constant. The equation of motion for
the $\psi $ field is:

\begin{equation}
2\mathcal{H}\psi ^{\prime }+\psi ^{\prime \prime }=\frac{2|\zeta |{{a}^{2}}}{%
\omega }\rho _{m}e^{-2\psi }  \label{psiddot}
\end{equation}

The conservation equations for the matter fields, $\rho _r$ and $\rho _m$
respectively, are: 
\begin{eqnarray}
\rho _m^{\prime}+3{\mathcal{H}} \rho _m &=&0 \\
\rho_r^{\prime}+4{\mathcal{H}} \rho_r^{\prime}&=&2\psi^{\prime}\rho_r.
\label{dotrho1}
\end{eqnarray}

\subsection{Linear Perturbations}

The perturbed components of the total energy-momentum ($T_{\mu \nu
}^{total}=T_{\mu \nu }^{mat}+T_{\mu \nu }^{\psi }+T_{\mu \nu }^{em}$) arise
from perturbations of the different matter fields which are time and space
dependent. In particular, we have $\rho _{m}\rightarrow \rho _{m}+\delta
\rho _{m}$, $\rho _{r}\rightarrow \rho _{r}+\delta \rho _{r}$ and $\psi
\rightarrow \psi +\delta \psi $. Note also that we have to perturb the fluid
4-velocity field, so we have $u_{i}\rightarrow u_{i}+\delta u_{i}$, where $i$
$=1,2,3$.

In order to have gauge-invariant equations we need to express the perturbed
energy-momentum tensor in terms of the gauge-invariant energy density,
pressure, scalar field and velocity field perturbations . The gauge
invariants $\delta \rho _{m}^{(gi)}$, $\delta \rho _{r}^{(gi)}$ and $\delta
\psi ^{(gi)}$ are defined in the same way as the gauge-invariant
perturbation of a general four-scalar, see equation (\ref{gauge}), so:

\[
\delta \rho _{m}^{(gi)}=\delta \rho _{m}+\rho _{m}^{\prime }(B-E^{\prime
}),\quad \delta p^{(gi)}=\delta p+p^{\prime }(B-E^{\prime }),\quad \delta
\psi ^{(gi)}=\delta \psi +\psi ^{\prime }(B-E^{\prime }), 
\]%
where $\delta p$ is the perturbed pressure for a specific component and the
gauge-invariant three-velocity $\delta u_{i}^{(gi)}$ is given by \cite%
{mukhanov}: 
\[
\delta u_{i}^{(gi)}=\delta u_{i}+a(B-E^{\prime })_{|i}. 
\]%
The physical meaning of the quantities which enter the gauge-invariant
equations is very simple: they coincide with the corresponding perturbations
in the longitudinal gauge. From now on we will drop the superscript $(gi)$
since we will always be dealing only with gauge-invariant quantities.

We can now write the gauge-invariant linearly perturbed energy-momentum
tensor:

\begin{eqnarray}
\delta T_{0}^{\mathit{matter}\ 0} &=&\delta \rho _{m},  \label{emmatter} \\
\delta T_{i}^{\mathit{matter}\ 0} &=&\left( \rho _{m}+p_{m}\right)
a^{-1}\delta u_{i}, \\
\delta T_{j}^{\mathit{matter}\ i} &=&-\delta p_{m}\delta _{j}^{i},
\end{eqnarray}

\begin{eqnarray}
\delta T_{0}^{\psi \ 0} &=&\omega a^{-2}\left( \psi ^{\prime }\delta \psi
^{\prime }-\psi ^{{\prime }^{2}}\Phi \right) , \\
\delta T_{i}^{\psi \ 0} &=&\omega a^{-2}\psi ^{\prime }\partial _{i}\delta
\psi , \\
\delta T_{j}^{\psi \ i} &=&\omega a^{-2}\left( \psi ^{{\prime }^{2}}\Phi
-\psi ^{\prime }\delta \psi ^{\prime }\right) \delta _{j}^{i},
\end{eqnarray}

\begin{eqnarray}
\delta T_{0}^{\mathit{em}\ 0} &=&\left( |\zeta |\delta \rho _{m}+\delta \rho
_{r}\right) e^{-2\psi }-2e^{-2\psi }\left( |\zeta |\rho _{m}+\rho
_{r}\right) \delta \psi ,  \label{emem} \\
\delta T_{i}^{\mathit{em}\ 0} &=&e^{-2\psi }\left( |\zeta |\rho _{m}+\rho
_{r}+p_{m}+p_{r}\right) a^{-1}\delta u_{i}, \\
\delta T_{j}^{\mathit{em}\ i} &=&\left( 2\delta \psi \left(
p_{m}+p_{r}\right) e^{-2\psi }-\left( \delta p_{m}+\delta p_{r}\right)
e^{-2\psi }\right) \delta _{j}^{i}.
\end{eqnarray}%
We have assumed there are no anisotropic stresses in the energy-momentum
tensors and we have considered only adiabatic perturbations; that is, we
consider the pressure perturbations to depend only on the energy-density
perturbations. In the dust and radiation cases this means $p_{m}=0$, $\delta
p_{m}=0$, or $p_{r}=\frac{1}{3}\rho _{r}$ and $\delta p_{r}=\frac{1}{3}%
\delta \rho _{r}$, respectively.

The fully perturbed gauge--invariant Einstein equations can be obtained from
(\ref{genpertein}) using the expressions above for $\delta T_{\nu }^{\mu }$ (%
\ref{emmatter}-\ref{emem}). We have

\begin{eqnarray}  \label{einspert00}
\Phi \left( -3\mathcal{H}^2 + 3k + 4G\pi \omega {\psi ^{\prime}}^2 \right) +
\triangledown^2 \Phi - 4G\pi \omega \psi ^{\prime}\delta\psi^{\prime}- 3%
\mathcal{H}\Phi^{\prime}=  \nonumber \\
4G\pi {a}^2 e^{-2\psi }\left[ \left( e^{2\psi } + |\zeta| \right)
\delta\rho_m + \delta\rho_r - 2\delta\psi \left( \rho_r + |\zeta|\rho_m
\right) \right]
\end{eqnarray}

\begin{equation}  \label{einspert01}
\mathcal{H} \triangledown^2 \Phi + \triangledown^2 \Phi^{\prime}= \\
4G\pi \omega \psi ^{\prime}\triangledown^2 \delta\psi -\frac{4G\pi}{3} a
e^{-2\psi } \left[ 4\rho_r + 3\left( e^{2\psi } + |\zeta| \right) \rho_m %
\right] \triangledown \delta u  \nonumber
\end{equation}

\begin{eqnarray}  \label{einspert11}
\Phi \left( {\mathcal{H}}^2 + 2\mathcal{H}^{\prime}- k + 4G\pi \omega {\psi
^{\prime}}^2 \right) + 3\mathcal{H}\Phi^{\prime}+ \Phi^{\prime\prime}= 
\nonumber \\
\frac{4G\pi}{3} \left[ {a}^2 e^{-2\psi}\left( \delta\rho_r -
2\rho_r\delta\psi \right) + 3\omega \psi ^{\prime}\delta\psi^{\prime}\right]
\end{eqnarray}

It is useful to write the perturbed energy-momentum conservation equations
for each component:

\begin{equation}  \label{psiperteq}
\delta \psi ^{^{\prime \prime }}=\frac{2|\zeta |}{\omega }e^{-2\psi }{a}^{2}%
\left[ \delta \rho _{m}+2\rho _{m}\left( \Phi -\delta \psi \right) \right] -2%
\mathcal{H}\delta \psi ^{^{\prime }}+\triangledown ^{2}\delta \psi +4\psi
^{\prime }\Phi ^{^{\prime }},
\end{equation}

\begin{equation}  \label{radpert}
\delta \rho _{r}^{^{\prime }}=-\frac{2}{3}\left[ \delta \rho _{r}\left( 6%
\mathcal{H}-3\psi ^{\prime }\right) +\rho _{r}\left( 2\triangledown \delta
u-3\delta \psi ^{^{\prime }}-6\Phi ^{^{\prime }}\right) \right] ,
\end{equation}

\begin{equation}  \label{dustpert}
\delta \rho _{m}^{^{\prime }}=-3\mathcal{H}\delta \rho _{m}-\rho _{m}\left(
\triangledown \delta u-3\Phi ^{^{\prime }}\right) .
\end{equation}

From these expressions it is clear that perturbations in $\alpha $ are
sourced by perturbations in the dust, but not vice versa. Hence we expect
that, in a dust-dominated universe with varying $\alpha $, the cold dark
matter perturbations will behave as in a dust-dominated universe with no
varying $\alpha $. Notice however that the same cannot be concluded so
easily for a radiation-dominated era, since there is a source term in
equation (\ref{radpert}) proportional to $\delta \psi ^{\prime }$. We expect
this term to be negligible at large scales, but that might not be the case
on small scales.

The gauge-invariant perturbation for $\alpha $ is given by eqn. (\ref{e}) as

\[
\frac{\delta \alpha }{\alpha }=2\delta \psi . 
\]%
From equations (\ref{einspert00}), (\ref{einspert01}), (\ref{einspert11}),
using (\ref{fried1}) to simplify, we obtain the general form for $\delta
\psi ,$ the perturbation to the scalar field which drives variations in the
fine structure 'constant', as 
\begin{eqnarray}
\delta \psi &=&\frac{1}{8G\pi {a}^{2}\left( 2\rho _{r}+3|\zeta |\rho
_{m}\right) }\left( 4G\pi {a}^{2}\left[ 2\left( \delta \rho _{r}-4\rho
_{r}\Phi \right) +3\left( e^{2\psi }+|\zeta |\right) \right. \right.
\label{psipert} \\
&&\left. \left. \left( \delta \rho _{m}-2\rho _{m}\Phi \right) \right]
+3e^{2\psi }\left[ 2\Phi \left( 3\mathcal{H}^{2}-k-4G\pi \omega {\psi
^{\prime }}^{2}\right) -\triangledown ^{2}\Phi +6\mathcal{H}\Phi ^{^{\prime
}}+\Phi ^{^{\prime \prime }}\right] \right)  \nonumber
\end{eqnarray}%
This is a gauge-invariant expression for $\delta \psi $, written as a
function of the gauge-invariant quantities $\Phi $, $\delta \rho _{m}$, and $%
\delta \rho _{r}$.

\section{Evolution of the Perturbations}

In a spatially flat universe, $\kappa =0$, the general gauge-invariant
expression for $\delta \psi $ becomes:

\begin{eqnarray}
\delta \psi &=&\frac{1}{8G\pi {a}^{2}\left( 2\rho _{r}+3|\zeta |\rho
_{m}\right) }\left( 4G\pi {a}^{2}\left[ 2\left( \delta \rho _{r}-4\rho
_{r}\Phi \right) +3\left( e^{2\psi }+|\zeta |\right) \right. \right.
\label{deltapsiflat} \\
&&\left. \left. \left( \delta \rho _{m}-2\rho _{m}\Phi \right) \right]
+3e^{2\psi }\left[ \Phi \left( 6\mathcal{H}^{2}-8G\pi \omega {\psi ^{\prime }%
}^{2}\right) -\triangledown ^{2}\Phi +6\mathcal{H}\Phi ^{^{\prime }}+\Phi
^{^{\prime \prime }}\right] \right) .  \nonumber
\end{eqnarray}

It was shown in \cite{bsm3} and \cite{Barrow:2002ed} that the cosmological
evolution of the metric scale factor is unchanged (up to very small
logarithmic corrections) to leading order by the time evolution of $\psi $.
The dominant effect is that of the evolution of the scale factor on the
evolution of $\psi $ through its propagation equation.

It is known that perturbations of massless scalar fields, or scalar fields
with a very small mass are negligible with respect to perturbations in the
matter fields and the gravitational potential. Guided by this, in order to
obtain the evolutionary behaviour for $\frac{\delta \alpha }{\alpha }$, we
will assume that the matter field perturbations, $\delta \rho _{m}$ and $%
\delta \rho _{r}$, and the metric perturbations, $\Phi $, are unaffected by
the $\psi $ perturbations to leading order. We will assume that these three
quantities will therefore be the same to this order as they are in a flat
FRW universe filled with barotropic matter and a minimally coupled scalar
field. These assumptions are valid if the energy density of $\psi $ is much
smaller than the energy density of the matter fields, so $\Phi $ will be
driven only by the matter perturbations. If we examine the perturbations in
the non-linear regime it is confirmed that $\frac{\delta \psi }{\psi }\ll 
\frac{\delta \rho }{\rho }$(see figure (\ref{dustdeltaalpha})).\textbf{\ }

\subsection{Radiation-Dominated Universes}

In any radiation-dominated era in which the expansion of the universe is
dominated by relativistic particles with an equation of state $p=\frac{1}{3}%
\rho $, we can neglect the non-relativistic stresses in the universe, in
particular, the cold dark matter and the cosmological constant since $\rho
_{r}>>\rho _{m}>>\rho _{\Lambda }$ to a good approximation. If we assume
that $\rho _{\Lambda }=\rho _{\Lambda }=0=\kappa $ and $\delta \rho _{m}=0$,
the background equations of motion give the usual conformal time evolution
for the scale factor and the energy density of the radiation: 
\[
\rho _{r}=\rho _{r_{0}}a^{-4}e^{2\psi }\qquad a=\sqrt{8\pi G\rho _{r_{0}}}%
\eta 
\]%
where $\rho _{r_{0}}$ is a constant.

The perturbations in the barotropic matter fluid and the potential $\Phi $
come from the equations (\ref{einspert00}), (\ref{einspert01}). The usual
flat FRW solutions with constant $\alpha $ are obtained by setting the terms
proportional to $\psi $ to zero, (\ref{einspert11}). The general solution of
these equations can be obtained by expanding the physical quantities in
terms of the eigenfunctions of the operator $\triangledown ^{2} $ (where $%
-k^{2}$ denotes the eigenvalue of this operator) and solving for each mode
separately. Resuming the terms, the general solutions of the linearly
perturbed equations for the potential $\Phi $ and the barotropic matter are:

\begin{equation}
\Phi =\eta ^{-3}\{\left[ w\eta \cos (w\eta )-\sin (w\eta )\right] C_{1}+%
\left[ w\eta \sin (w\eta )+\cos (w\eta )\right] C_{2}\}e^{i\mathbf{kx}}
\label{radbackgroundphi}
\end{equation}%
and 
\begin{eqnarray}
\frac{{\delta \rho _{r}}}{\rho _{r}} &=&\frac{4}{\eta ^{3}}\left(
C_{1}\{\eta w\left[ 1-\frac{1}{2}\left( \eta w\right) ^{2}\right] \cos (\eta
w)+\left[ \left( \eta w\right) ^{2}-1\right] \sin (\eta w)\}\right.
\label{radbackgroundrho} \\
&&+\{\left[ 1-\left( \eta w\right) ^{2}\right] \cos (\eta w)+\eta w\left[ 1-%
\frac{1}{2}\left( \eta w\right) ^{2}\right] \sin (\eta w)\}C_{2})e^{i\mathbf{%
kx}},  \nonumber
\end{eqnarray}%
where we have expanded the general solution in plane waves since we are
assuming a spatially flat universe; $C_{1}$ and $C_{2}$ are arbitrary
functions of the spatial coordinates; $k$ is the wave vector mode and $w=k/%
\sqrt{3}$. Note that these quantities are all expressed in their
gauge-invariant format. Finally, to calculate the explicit time dependence
of $\delta \psi $, we need to use the background solution for $\psi $:%
\begin{equation}
\psi =\frac{1}{2}\log {(8N)}+\frac{1}{4}\log {(\frac{a_{0}}{2}\eta ^{2}).}
\label{radol}
\end{equation}%
This was found in \cite{bsm2} and \cite{Barrow:2002ed} for a
radiation-dominated universe, where $N=-\frac{2\zeta }{\omega }\rho
_{m}a^{3}>0$ since $\zeta <0$ in the magnetic energy dominated theories
considered by BSBM. It is important to notice that in universes with an
entropy to baryon ratio ($S\sim 10^{9}$) like our own, $\psi $ does not
experience any growth in time \cite{bsm1}. The constant term on the
right-hand side of (\ref{radol}) dominates the solution for $\psi (\eta )$
throughout the radiation era. Numerical solutions confirm this freezing in
of $\psi $, and hence of $\alpha $, during the radiation era,\ \cite{bsbm}.

\subsubsection{Large-scale perturbations in a radiation-dominated era}

In the long-wavelength limit ($w\eta <<1$) where the scale of the
perturbation exceeds the Hubble radius, we can neglect spatial gradients.
So, in this limit, $\frac{\delta \alpha }{\alpha }$ becomes:

\[
\delta \psi =\frac{1}{2}\frac{\delta \alpha }{\alpha }\propto \frac{1}{2\eta 
}e^{i\mathbf{kx}}\triangledown ^{2}C_{2}-4\pi Gwe^{i\mathbf{kx}}C_{2}\eta
^{-3} 
\]

From this expression we can see at that on large scales the inhomogeneous
perturbations in $\alpha $ will decrease as a power-law in time. This
behaviour agrees with the one found in \cite{bsm3} by other methods.

\subsubsection{Small-scale perturbations in a radiation-dominated era}

On scales smaller than the Hubble radius ($w\eta >>1$) the dominant terms
are proportional to $\triangledown ^{2}C_{1}$ and $\triangledown ^{2}C_{2}$,
so the asymptotic behaviour for $\frac{\delta \alpha }{\alpha }$ will be: 
\emph{\ }

\begin{eqnarray}
\delta \psi &=&\frac{1}{2}\frac{\delta \alpha }{\alpha }\propto -\frac{1}{2}w%
\left[ \cos (\eta w)\triangledown ^{2}C_{1}+\sin (\eta w)\triangledown
^{2}C_{2}\right] e^{i\mathbf{kx}}+  \label{deltaradsmall} \\
&&+\frac{1}{2\eta }\left[ \sin (\eta w)\triangledown ^{2}C_{1}-\cos (\eta
w)\triangledown ^{2}C_{2}\right] e^{i\mathbf{kx}}  \nonumber
\end{eqnarray}

On small scales we can see the perturbations on $\alpha $ will be
oscillatory. This behaviour is new and does not coincide with the ones found
in \cite{bsm3}.

\subsection{Dust-Dominated Universes}

In the case of a flat dust-dominated universe, filled with a $p=0$ fluid, we
can assume $\rho _{r}=\rho _{\Lambda }=0=\kappa $ and $\delta \rho _{r}=0$.
Again, in order to obtain an explicit expression of $\delta \psi $ from
equation (\ref{deltapsiflat}), we will assume that the matter-field
perturbations, $\rho _{m}$ and $\delta \rho _{m}$, and the metric
perturbation, $\Phi $, are not affected by the $\psi $ perturbations to
leading order. Thus, we will assume that these functions behave as in a
perturbed flat FRW dust universe.

In general, for a flat dust universe we have the following background
solutions:

\[
\rho _{m}=\rho _{m_{0}}a^{-3},\qquad a=\frac{2\pi G\rho _{m_{0}}}{3}\eta
^{2}, 
\]%
where $\rho _{m_{0}}$ is constant.

As before, we can calculate the most general gauge-invariant solutions for
the energy-momentum perturbations $\delta \rho _{m},$ and for the potential $%
\Phi $, assuming that the $\psi $ field does not affect them to leading
order, so their time dependences are \cite{mukhanov}:

\begin{equation}
\Phi =C_{1}+C_{2}\eta ^{-5}  \label{dustbackground}
\end{equation}%
and 
\begin{equation}
\frac{\delta \rho _{m}}{\rho _{m}}=\frac{1}{6}\left[ \left( \eta
^{2}\triangledown ^{2}C_{1}-12C_{1}\right) +\left( \eta ^{2}\triangledown
^{2}C_{2}+18C_{2}\right) \eta ^{-5}\right] ,
\end{equation}%
where $C_{1}$ and $C_{2}$ are arbitrary functions of the spatial coordinates.

Once again, we need the background solution for $\psi $, in order to
calculate $\delta \psi $ as an explicit function of the conformal time. We
use the asymptotic solution 
\[
\psi =\frac{1}{2}\log {[2N\log {(\frac{a_{m}}{6}\eta ^{3})}],} 
\]%
which was found in \cite{bsm3}, as an asymptotic approximation for a
dust-dominated universe which is in good agreement with numerical solutions,
and where, as above, $N=-\frac{2\zeta }{\omega }\rho _{m}a^{3}>0$ is a
constant.

\subsubsection{Large-scale perturbations in a dust-dominated era}

On scales larger than the Hubble radius we can neglect the terms
proportional to $\partial _{i}C_{1}$,$\triangledown ^{2}C_{1}$, $%
\triangledown ^{2}C_{2}$ and $\partial _{i}C_{2}$; so in this limit we have
the following asymptotic behaviour for the non-decaying mode:

\begin{equation}
\delta \psi =\frac{1}{2}\frac{\delta \alpha }{\alpha }\propto -2C_{1}-\frac{%
2\pi G\rho _{m_{0}}}{\ln (\eta )}C_{1}  \label{deltadustlarge}
\end{equation}%
Therefore, on large scales the inhomogeneous perturbations of $\alpha $ will
not grow in time by gravitational instability. This behaviour can be
understood with reference to the general evolution equation for $\psi $ in
Friedmann universes. On large scales, where spatial gradients in $\psi $ can
be neglected with respect to its time derivatives, we may view
inhomogeneities in density and in $\psi $ as if they are separate Friedmann
universes of non-zero curvature ($\kappa \neq 0$). The growth of
inhomogeneity can be deduced by comparing the evolution of $\alpha $ in the $%
\kappa \neq 0$ universes with those in the $\kappa =0$ model (a more
detailed numerical study of this model will be presented in section 5
below). In effect, this uses the Birkhoff-Newton property of gravitational
fields with spherical symmetry. We note that the (\ref{psiddot}) evolution
equation has the simple property that $\psi $ cannot have a maximum because $%
\ddot{\psi}>0$ when $\dot{\psi}>0$ because $N>0$ , \cite{bsm3}. This result
holds irrespective of the value of $\kappa \lesseqqgtr 0.$ Thus $\psi $ and $%
\alpha $ will continue their slow increase in both over-densities,
under-densities and the flat background until we reach scales small enough
for spatial derivative to come significantly into play. This has the
important consequence that we do not expect large spatial inhomogeneities in 
$\alpha $ to have developed. However, it should be noted that the
sensitivity of the observations of varying-$\alpha $ effects in quasar
spectra is sufficient to discern variations in redshift space smaller than $%
O(10^{-5})$, which is of the same order as the amplitude of density
fluctuation on very large scales in the universe. \emph{\ }

\subsubsection{Small-scale perturbations in a dust-dominated era}

On scales smaller than the Hubble radius the dominant terms are those
proportional to $\triangledown ^{2}C_{1}$ and $\triangledown ^{2}C_{2}$, so
the asymptotic behaviour for the growing mode will be:

\begin{equation}
\delta \psi \propto \frac{1}{12}\eta ^{2}\triangledown ^{2}C_{1}
\label{deltadustsmall}
\end{equation}%
This also shows that perturbations of $\alpha $ will grow on small scales.
This result is a product of the assumption that on small scales the universe
can be considered as being filled by an homogeneous and isotropic fluid,
however we know this is not true below the scale where gravitational
clustering becomes non-linear. On these small scales we also have to worry
about new consequences of inhomogeneity which have not been included in our
analysis. For example, the constant parameter $N=-\frac{2\zeta }{\omega }%
\rho _{m}a^{3}\varpropto \zeta \Omega _{m}$ will vary in space due to
inhomogeneity in the background matter density parameter $\Omega _{m}$ and
in the dark matter parameter $-1\leq \zeta \leq 1$. We have assumed that $%
\zeta <0$ for the cold dark matter on large scales in order for the
cosmological consequences of time-varying $\alpha $ to be a small
perturbation to the standard cosmological dynamics. But on small scales the
dark matter will be baryonic in nature and so $\zeta >0$ there. Hence, we
expect $\zeta $ to be significantly scale dependent as we go to small
scales. This behaviour will be investigated elsewhere along with the problem
of the clustering of inhomogeneities in $\rho $ and $\alpha $.

\subsection{Accelerated Expansion}

In an era of accelerated expansion, $\ddot{a}>0$, as would arise during
inflation or during a $\Lambda $- or quintessence-dominated epoch at late
times, we can consider the scale factor to evolve as a power-law of the
conformal time as $a=\eta ^{-n}$, where $n\geq 1$ and $\eta $ \ runs from $%
-\infty $ to $0$. The case of $a=\eta ^{-1}$ corresponds to a $\Lambda $%
-dominated epoch.

As in the previous sections, we will assume that all the other matter
components which fill the universe will behave exactly as in a perturbed FRW
universe with no variations in $\alpha $. We also assume that neither of the
dust and radiation perturbations will affect the behaviour and evolution of $%
\delta \psi $, and that these perturbations are negligible with respect to
the $\Lambda $ stress driving the expansion, so we will consider $\delta
\rho _{m}=0$ and $\delta \rho _{r}=0$.

It was found in \cite{bsbm} and \cite{Barrow:2002ed} that in universe which
is undergoing accelerated expansion, the asymptotic solution for $\psi $ is
a constant, so we will assume that $\psi =\psi _{\infty }$, in the
background, where $\psi _{\infty }$ is a constant. Thus, equations (\ref%
{einspert00}) and (\ref{einspert11}) become: 
\[
\frac{8G\pi |\zeta |\delta \psi \rho _{m}}{e^{2\psi _{\infty }}\eta ^{2n}}%
+\triangledown ^{2}\Phi +\frac{3n\left( \eta \Phi -n\Phi ^{^{\prime
}}\right) }{\eta ^{2}}=0 
\]%
\[
n\left( 2+n\right) \Phi +\eta \left( \eta \Phi ^{^{\prime \prime }}-3n\Phi
^{^{\prime }}\right) =0 
\]%
where we have also considered $\rho _{r}=0$, but $\rho _{m}\neq 0$ because
of the coupling with $\psi $ in the equation of motion of the scalar field.
Note that if we had also set $\rho _{m}=0$ here, we would have imposed a no $%
\alpha $-variation condition: $\delta \psi =0$.

Integrating the last equation, we obtain

\[
\Phi =\eta ^{\frac{3n-\sqrt{1+n\left( 5n-2\right) }}{2}}\left( \sqrt{\eta }%
C_{1}+\eta ^{\frac{1}{2}+\sqrt{1+n\left( 5n-2\right) }}C_{2}\right) 
\]%
where $C_{1}$ and $C_{2}$ are arbitrary functions of the spatial
coordinates. Note that since $n>1$ in accelerating universes we see that $%
\Phi $ will \textbf{decay in time as $\eta\rightarrow0$}. From this solution
for $\Phi $, we obtain:

\begin{eqnarray}
\delta \psi  &=&\frac{1}{2}\frac{\delta \alpha }{\alpha }=-\frac{e^{2\psi
_{\infty }}}{16G\rho _{m_{0}}\pi |\zeta |}\eta ^{\frac{-3+n-\sqrt{1+n\left(
-2+5n\right) }}{2}}\left[ 3n\left( 1+n-\sqrt{1+n\left( 5n-2\right) }\right)
C_{1}+\right.  \\
&&\left. 3n\left( 1+n+\sqrt{1+n\left( 5n-2\right) }\right) \eta ^{\sqrt{%
1+n\left( 5n-2\right) }}C_{2}+2\eta ^{2}\triangledown ^{2}C_{1}+2\eta ^{2+%
\sqrt{1+n\left( 5n-2\right) }}\triangledown ^{2}C_{2}\right]   \nonumber
\end{eqnarray}%
which is a decaying function of the conformal time when $n>1$ and a constant
when $n=1$. Thus, in accord with the expectations of the cosmic no hair
theorem, the universe approaches the FRW model and $\alpha $ is
asymptotically constant at late times.

\subsubsection{Large-scale perturbations during accelerated expansion}

On scales larger than the Hubble radius we can neglect the terms
proportional to the spatial derivatives; so in this limit we have the
following asymptotic behaviour:

\begin{eqnarray}
\delta \psi  &\propto &-\frac{3e^{2\psi _{\infty }}n}{16G\rho _{m_{0}}\pi
|\zeta |}\eta ^{\frac{n-3-\sqrt{1+n\left( 5n-2\right) }}{2}}\left[ \left(
1+n-\sqrt{1+n\left( 5n-2\right) }\right) C_{1}\right.   \nonumber \\
&&\left. +\left( 1+n+\sqrt{1+n\left( 5n-2\right) }\right) \eta ^{\sqrt{%
1+n\left( 5n-2\right) }}C_{2}\right] 
\end{eqnarray}%
Therefore, as expected, on large scales during an accelerated era
inhomogeneities in $\alpha $ will decrease on time when $n>1$ and will be a
constant when $n=1$.

\subsubsection{Small-scale perturbations during accelerated expansion}

On scales smaller than the Hubble radius the dominant terms are the ones
proportional to $\triangledown ^{2}C_{1}$ and $\triangledown ^{2}C_{2}$, so
the asymptotic behaviour will be

\begin{equation}
\delta \psi \propto -\frac{e^{2\psi _{\infty }}}{8G\rho _{m_{0}}\pi |\zeta |}%
\eta ^{\frac{1+n-\sqrt{1+n\left( 5n-2\right) }}{2}}\left[ \triangledown
^{2}C_{1}+\eta ^{\sqrt{1+n\left( 5n-2\right) }}\triangledown ^{2}C_{2}\right]
\end{equation}%
This confirms that perturbations of $\alpha $, as $\eta \rightarrow 0$, will
decrease on small scales when $n>1$ and will be constant when $n=1$.

\subsection{\protect\bigskip Summary of behaviour}

\vspace{.5cm}

In Table 1 we summarise the time evolution of small inhomogeneities in $%
\alpha $\ found under different conditions in this section.

\begin{center}
\begin{tabular}{|c|c|c|}
\hline
\textbf{Universal equation of state} & \multicolumn{2}{|c|}{\textbf{Time} 
\textbf{Evolution of the perturbations $\delta \psi =$}$\frac{1}{2}\frac{%
\delta \alpha }{\alpha }$} \\ \cline{2-3}
& \textit{Large scales} & \textit{Small scales} \\ \hline
\textit{Radiation-dominated epoch} & Decaying & Oscillatory \\ 
$p=\frac{1}{3}\rho ,\qquad a\propto \eta $ &  &  \\ \hline
\textit{Dust-dominated epoch} & Constant & Growing \\ 
$p=0,\qquad a\propto \eta ^{2}$ &  &  \\ \hline
\textit{Accelerated expn }$\mathit{{a\propto \eta ^{-n},n\geq 1}}$\textit{\ }
&  &  \\ 
$\Lambda $-dominated & Constant & Constant \\ 
$p=-\rho ,\qquad a\propto \eta ^{-1}$ &  &  \\ \cline{2-3}
Power-law acceleration, $n>1$ & Decaying & Decaying \\ 
$p=w\rho ,\qquad w<0,\qquad a\propto \eta ^{-n}$ &  &  \\ \hline
\end{tabular}
Table 1: Time evolution of small inhomogeneities in $\alpha .$
\end{center}

\vspace{.5cm}

\section{The Non-Linear Regime}

In order to study the evolution of inhomogeneities in $\alpha $ beyond the
domain of linear perturbation theory we need to use a different model. The
simplest approach is to confine attention to spherically symmetric
inhomogeneities. This will be done by comparing the solution of the BSBM
theory for $\alpha $ in a closed ($\kappa =1$) universe, with the solution
for $\alpha $ in a flat ($\kappa =0$) universe. We are assuming a Birkhoff
property for the BSBM theory so that we can treat the perturbation as an
independent closed universe. This is a standard technique in general
relativity which was first used by Lema\^{\i}tre \cite{lem}.

We define the alpha 'over-density perturbation' (which is not necessarily
small) by

\[
\frac{\delta \alpha }{\alpha }\equiv \frac{\alpha _{\kappa =1}-\alpha
_{\kappa =0}}{\alpha _{\kappa =0}} 
\]%
where $\alpha _{\kappa }$ is the solution of equation (\ref{psieq}) for a
universe with curvature $\kappa $.

\subsection{Radiation-dominated Era}

\begin{figure}[t]
\epsfig{file=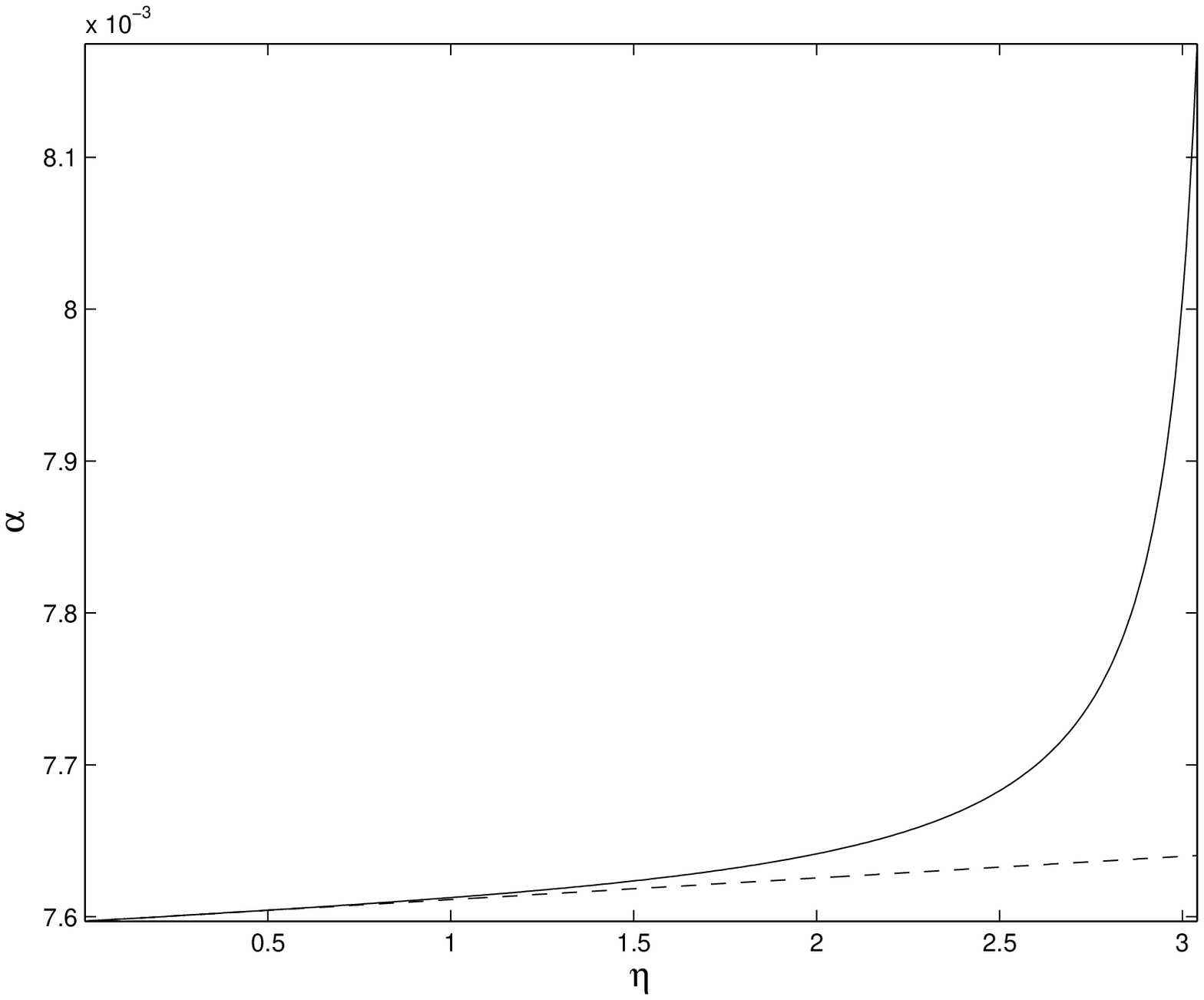,height=4.68cm} \epsfig{file=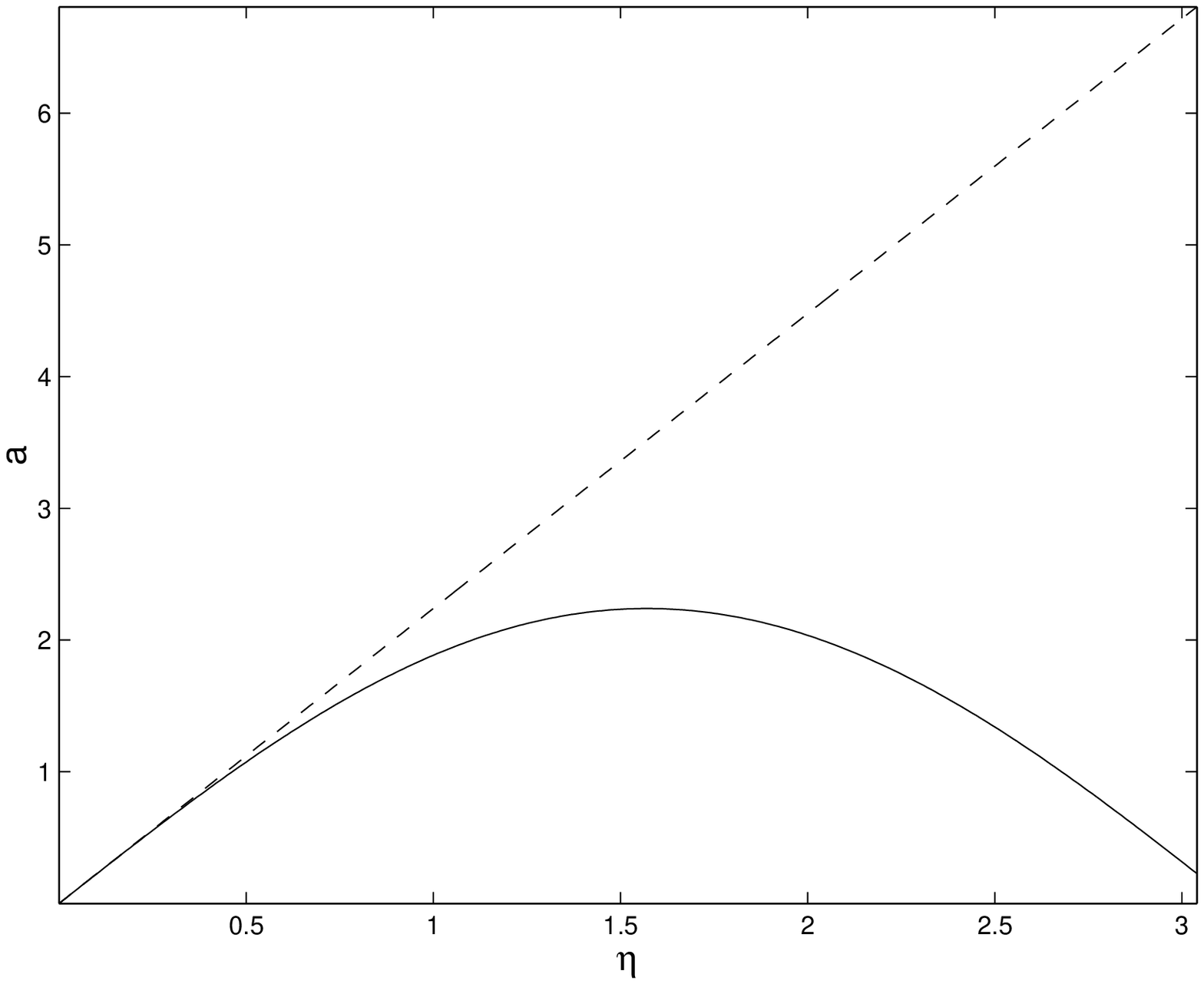,height=4.68cm}
\caption{{\protect\small \textit{\ The evolution of $\protect\alpha(\protect%
\eta)$ and $a(\protect\eta)$ for radiation-dominated universes with $\protect%
\kappa=0$ (dashed) and $\protect\kappa=1$ (solid).}}}
\label{radalpha}
\end{figure}

\begin{figure}[t]
\epsfig{file=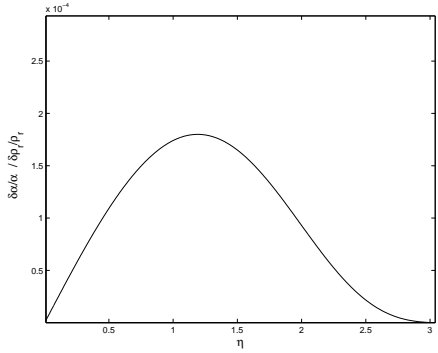,height=4.68cm}
\caption{{\protect\small \textit{The evolution of $\frac{\protect\delta%
\protect\alpha}{\protect\alpha}/ \frac{\protect\delta\protect\rho_r}{\protect%
\rho_r}$ vs $\protect\eta$ in radiation-dominated universes. }}}
\label{raddeltaalpha}
\end{figure}

The scale factor for a radiation-dominated closed ($\kappa =1$) FRW universe
is given by $a=\sin (\eta )$; for a flat ($\kappa =0$) FRW universe the
normalised scale factor is given by $a=\eta $.

The evolution of $\alpha $ can be seen in Figure \ref{radalpha} along with
the evolution of $a$ for $\kappa =1,0$. As expected, we can see there is no
difference in the evolution of $\alpha $ at early times and $\alpha \propto
\eta $ as it was found in \cite{Barrow:2002ed}. When the difference between
the scale factors of the two universes becomes significant, the behaviour of 
$\alpha _{\kappa =1}$ begins to deviate from that of $\alpha _{\kappa =0}$.
We see that $\alpha _{\kappa =1}$ clearly grows faster than $\alpha _{\kappa
=0}$. The difference in the growth rates become very significant near the
expansion maximum of the bound region ($\eta \approx \pi $). However, after
this time our assumption that the background is not affected by changes of $%
\psi $ in the cosmological equations that describe the background universe
breaks down, since the kinetic energy of the scalar field will diverge and
can no longer be neglected in the Friedmann equation. We expect the
behaviour near the final singularity to be similar to the kinetic-dominated
evolution near the initial singularity discussed in ref. \cite{bsm1}.

From Figure \ref{raddeltaalpha} we see that the variations in $\alpha $ will
become increasingly important as $\eta $ approaches the second singularity.
Notice that although there is a considerable growth in the perturbations in $%
\alpha $, when we compare them with the perturbations in the radiation, they
are not as significant as can be observed from the evolution of the ratio $%
\frac{\delta \alpha }{\alpha }/\frac{\delta \rho _{r}}{\rho _{r}}$ versus $%
\eta $ in Figure \ref{raddeltaalpha}.

\subsection{Dust-dominated Era}

\begin{figure}[t]
\epsfig{file=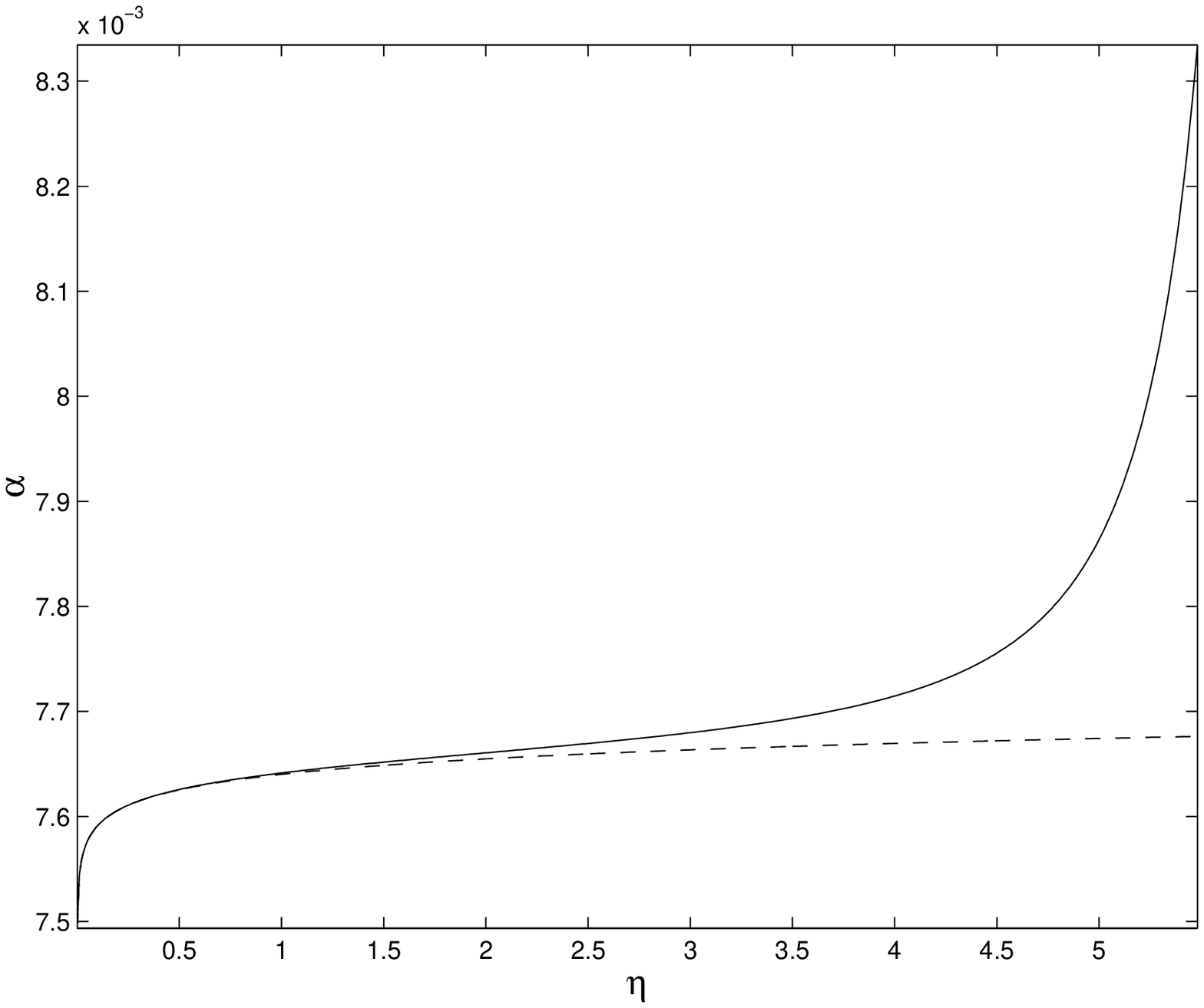,height=4.68cm} %
\epsfig{file=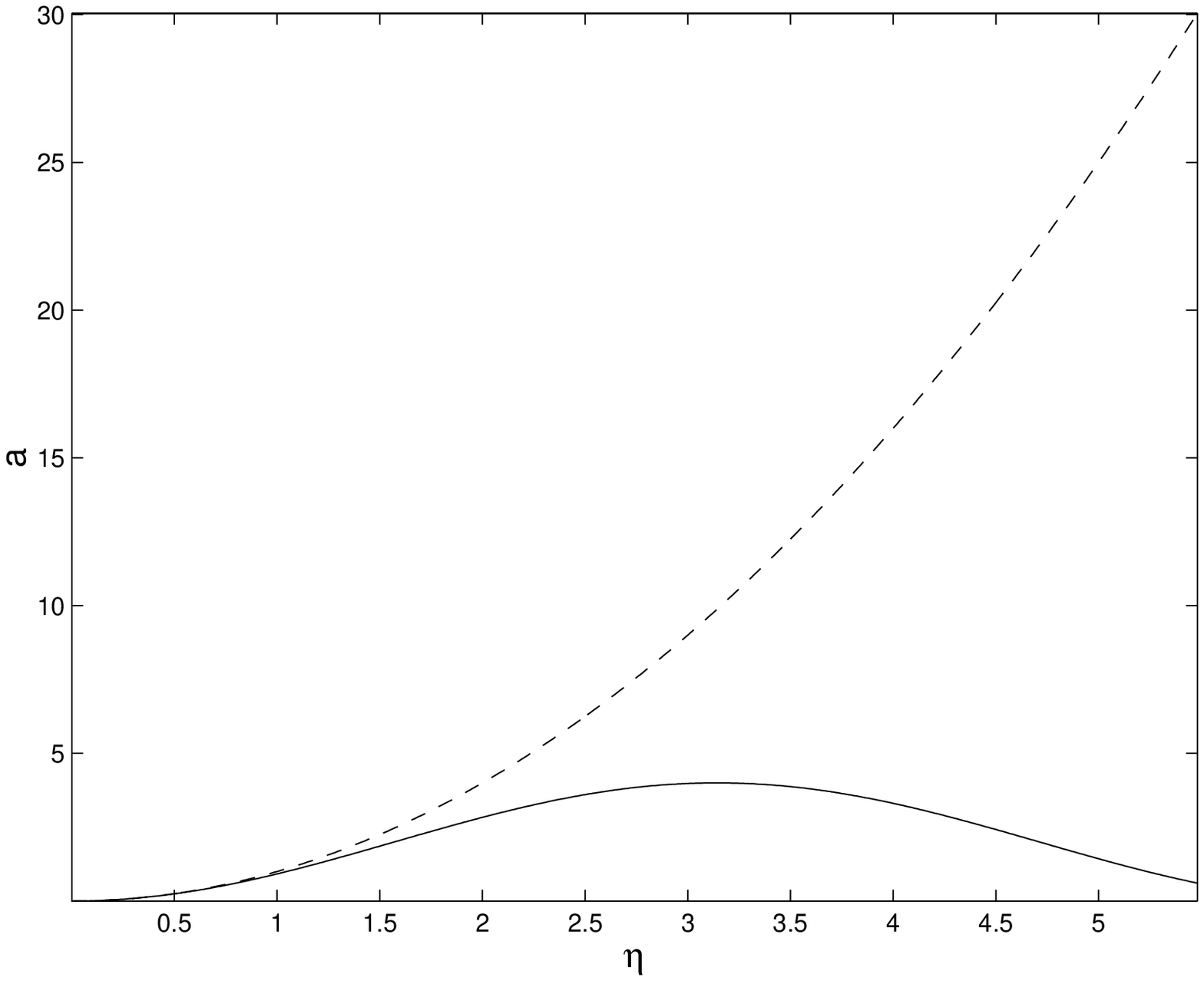,height=4.68cm}
\caption{{\protect\small \textit{The evolution of $\protect\alpha(\protect%
\eta)$ and $a(\protect\eta)$ for dust-dominated universes with $\protect%
\kappa=0$ (dashed) and $\protect\kappa=1$ (solid). }}}
\label{dustalpha}
\end{figure}

\begin{figure}[t]
\epsfig{file=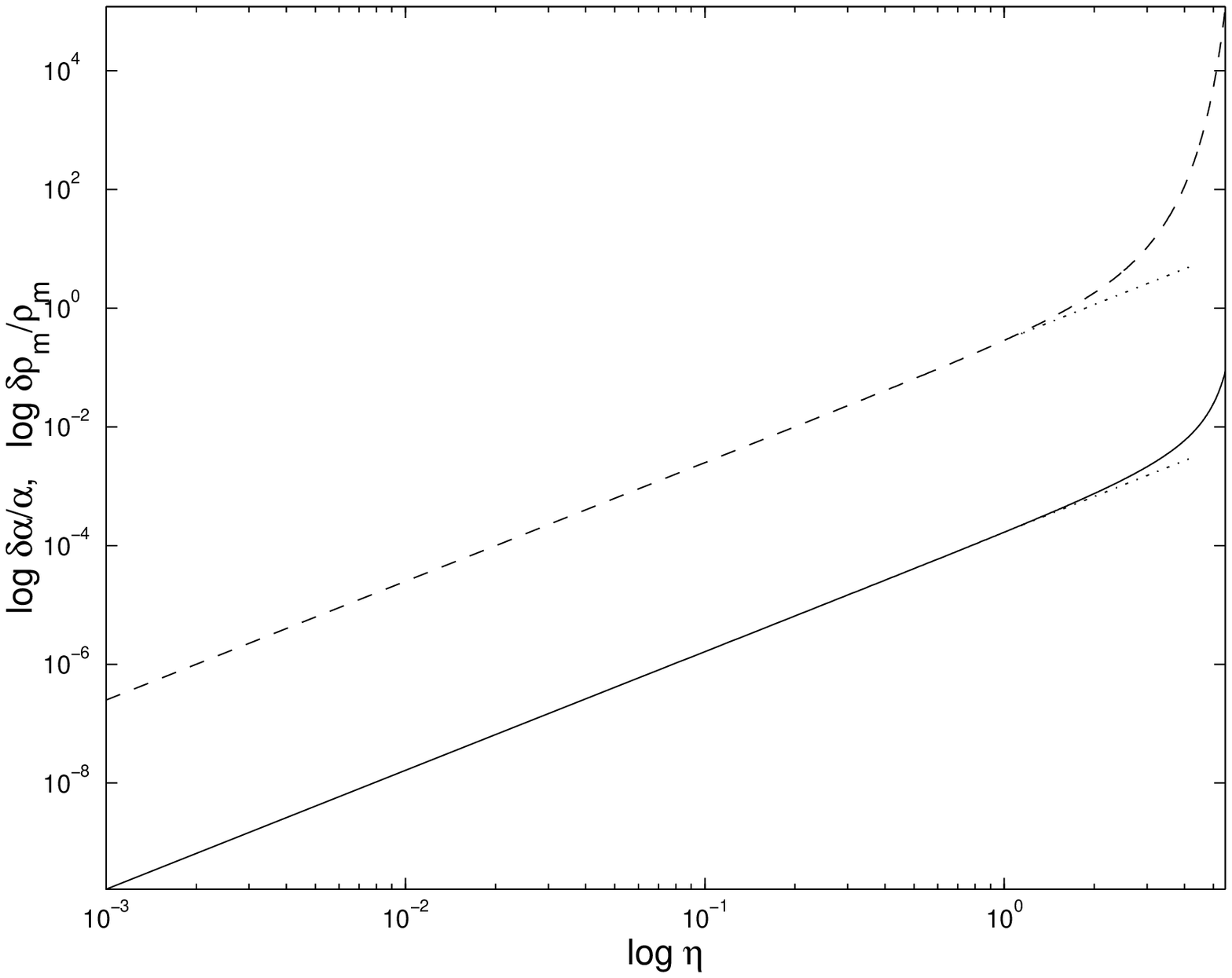,height=4.68cm} %
\epsfig{file=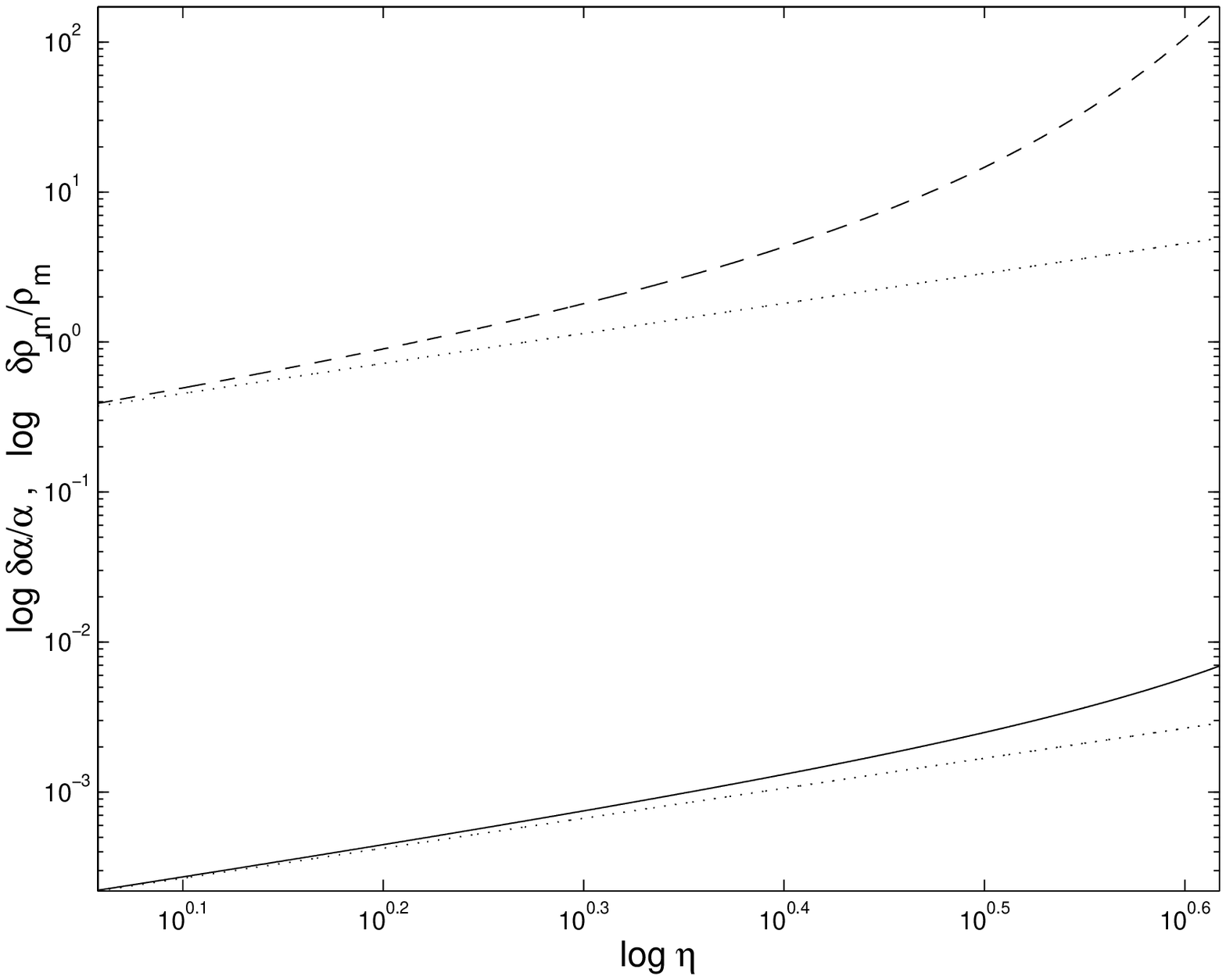,height=4.68cm}
\caption{{\protect\small \textit{The evolution of $\log \frac{\protect\delta%
\protect\alpha}{\protect\alpha}$ (solid) and $\log \frac{\protect\delta%
\protect\rho_m}{\protect\rho_m}$ (dashed) vs $\log(\protect\eta)$ for a
dust-dominated universe. The dotted lines correspond to evolution in the
linear perturbation solution. }}}
\label{dustdeltaalpha}
\end{figure}

During the dust-dominated phase of a closed universe ($\kappa =1$) the
normalised scale factor is given by $a=2\left( 1-\cos (\eta )\right) $,
while for a flat universe ($\kappa =0$) the normalised scale factor is given
by $a=\ \eta ^{2}$. Integrating equation (\ref{psieq}) for both cases we
obtain the evolution of $\alpha $ for both cases.

The evolution of $\alpha (\eta )$ can be seen in Figure \ref{dustalpha}
along with the evolution of $a(\eta )$ for $\kappa =1,0$. As in the
radiation case, we can see there is little difference in the evolution of $%
\alpha $ at early times, since the scale factor for the closed model evolves
very similarly to the flat one for $\eta <<1$ and $\alpha \propto \ln (\eta
) $. The differences between $\alpha _{\kappa =1}$ and $\alpha _{\kappa =0}$
cases start to appear when $\eta \approx 1$, when the nonlinear regime
commences. These differences become more accentuated near the second
singularity, but once again this is the region where our approximations
break down.

Notice that although there is a considerable growth in the perturbations in $%
\alpha $, when we compare them with the perturbations in the cold dark
matter, they are not as significant as can be observed from the evolution of 
$\log (\frac{\delta \alpha }{\alpha })$ and $\log (\frac{\delta \rho _{m}}{%
\rho _{m}})$ $vs.$ $\log (\eta )$ in Figure \ref{dustdeltaalpha}. In this
case perturbations in $\alpha $ are even less significant than in the
radiation case. We note that the fact that the linear regime is a very good
approximation at early times, since as can be seen from Figure \ref%
{dustdeltaalpha}, $\frac{\delta \alpha }{\alpha }$ tracks $\frac{\delta \rho
_{m}}{\rho _{m}}\propto t^{2/3}$ at early times. From the detail of Figure %
\ref{dustdeltaalpha} we see that variations in the cold dark matter will
start to occur before than variations in $\alpha $, and $\frac{\delta \rho
_{m}}{\rho _{m}}$ is at least three orders of magnitude bigger than $\frac{%
\delta \alpha }{\alpha }$.

\section{Conclusions}

By applying the gauge-invariant formalism of ref. \cite{mukhanov} to the
simple BSBM theory of varying $\alpha $ we have determined the evolution of
small inhomogeneities in $\delta \alpha /\alpha $ in the presence of small
adiabatic density inhomogeneities. To leading order, the evolution of the
perturbations to the expansion dynamics and the matter and radiation content
of the universe behave as in cosmological models with constant $\alpha $ and
we can to determine the behaviour of small inhomogeneities in $\delta \alpha
/\alpha $ in the gravitational fields created by the density and metric
perturbations (see also refs. \cite{bsm3} and \cite{Barrow:2002ed}).

In a flat radiation-dominated universe we find that inhomogeneous
perturbations in $\alpha $ will decrease on large scales while on scales
smaller than the Hubble radius they will undergo bounded oscillations. In
reality, we expect dissipation of the adiabatic fluctuations to occur by
Silk damping and small-scale fluctuations in $\alpha $ will also undergo
decay as their driving terms damp out. However, while the exact solution for
the evolution of $\alpha $ is a linear sum of a constant and a slow
power-law growth the power-law evolution does not become dominant by the end
of the radiation era in universes like our own with entropy per baryon $%
O(10^{9}).$

In a flat dust-dominated universe small inhomogeneities in $\alpha $ will
become constant on large scales at late times while on small scales they
will increase as $t^{2/3}$. In reality, the small-scale evolution will be
made more complicated by the breakdown of the assumptions underlying the
perturbation analysis and the development of local deviations from the FRW
behaviour.

In an accelerated phase of our universe, as is the case for an early
inflationary epoch, or during a $\Lambda $- or quintessence-dominated late
phase of evolution, we show that inhomogeneous perturbations in $\alpha $
will decrease on all scales. This result complements the earlier discovery 
\cite{bsbm}, \cite{bsm1}, that $\alpha $ tends to a constant with
exponential rapidity in Friedmann universes that become dominated by a
cosmological vacuum stress. Any pre-existing inhomogeneities will be frozen
in but their scale will be exponentially increased by the de Sitter
expansion.

These perturbative results are quite good approximations when we consider
large scales, but are expected to break down when extended to small scales,
where non-linear effects come into play and local deviations from isotropy
and homogeneity of the matter content are significant. We note that the
background solutions for $\psi $, about which we have linearised the
perturbations of the Einstein equations, are solutions which describe the
time evolution of $\psi $ on a 'standard' FRW background. Our neglect of the
back-reaction of the $\psi $ perturbations on the background expansion
dynamics is a good approximation up to logarithmic corrections.

When we examined the non-linear evolution of spherical inhomogeneities by
means of a comparative numerical study of flat and closed Friedmann models
we found that perturbations in $\alpha $ remain almost negligible with
respect to perturbations in the fluid that dominates the energy density of
the universe at the same epoch. Comparing the flat with the closed solution
for $\alpha $, we concluded that in both the cases of a radiation or
dust-dominated epoch, $\alpha $ will 'feel' the change of behaviour in the
scale factor, and the perturbations $\frac{\delta \alpha }{\alpha }$ will
grow in time. The early-time behaviour of the non-linear solutions confirms
the linear behaviour found in section 4. In particular$\frac{\delta \alpha }{%
\alpha }$ changes in proportion to $\frac{\delta \rho }{\rho }$ at early
times.

We have provided a detailed analysis of the behaviour of inhomogeneous
perturbations in $\alpha $ and its time variation on large scales under the
assumption that the defining constant of the BSBM theory, $\zeta $, is
constant and negative in sign. In reality this assumption will break down on
small scales. The negativity of the effective value of $\zeta $ requires
that the cold dark matter is dominated by the magnetic rather than the
electric field energy (see also the discussion of ref \cite{bsbm} and by
Bekenstein \cite{bek3}). However, on sufficiently small scales the dark
matter will become dominated by baryons and the sign of the effective $\zeta 
$ will have to change sign. Overall, there will also be a gradient in the
value of $\left\vert \zeta \right\vert $ reflecting the scale dependence of
the relative contribution of dark matter to the total density of the
universe. We have not included these effects in the present analysis. They
would need \ to be included in any detailed analysis of the small-scale
behaviour of inhomogeneities in $\alpha .$ This is an important challenge
for future work because it would enable the quasar data on varying $\alpha $
to be compared directly with the limits from the Oklo natural reactor \cite%
{fuj, dd} and Rhenium-Osmium abundances in meteorites \cite{dys, olive1}. At
present the relation between the cosmological and geonuclear evidences is
unclear because the latter are derived from physical processes occurring
within the cosmologically non-evolving solar system environment.

Variations in $\alpha $ also affect the cosmic microwave background
radiation spectrum and anisotropy in different ways, but the effects must be
disentangled from allowed changes in other cosmological parameters that can
contribute similar effects. These changes in the microwave background with $%
\alpha $ left as a free constant parameter were analysed in \cite{martins}
using the new WMAP data. They are far less sensitive that the many-multiplet
analyses of quasars at $z=0.5-3$\ \cite{murphy,webb2,webb,webb3}, although
they derive from higher redshifts, $z<1100$. These studies can accommodate
constant and varying $\alpha $ but up to a level that would be too large to
be consistent with the quasar data and the slow time-evolution of the theory
with time-varying $\alpha $\ described in this paper.

\vspace{2cm} \noindent \textbf{Acknowledgements }DFM would like to thank P.
Ferreira, K. Moodley and C. Skordis for hospitality at the Oxford University
Physics Department during the time when part of this article was written and
for discussions. DFM was partially supported by Funda\c{c}\~{a}o para a Ci%
\^{e}ncia e a Tecnologia, through the research grant BD/15981/98 and by Funda%
\c{c}\~{a}o Calouste Gulb\^{e}nkian, through the research grant Proc.50096.
We would like to thank J. Magueijo, M. Murphy, H. Sandvik, and J. Webb for
discussions.

\end{document}